# Simulations of Disordered Matter in 3D with the Morphological Autoregressive Protocol (MAP) and Convolutional Neural Networks

Ata Madanchi[1], Michael Kilgour[2,$], Frederik Zysk[3], Thomas D. Kühne[3], Lena Simine[2*]

**Affiliations:**

[1]Department of Physics, McGill University, 3600 University St., Montreal, Quebec, H3A 2T8, Canada.

[2]Department of Chemistry, McGill University, 801 Sherbrooke St. W, Montreal, Quebec, H3A 0B8, Canada.

[3]Dynamics of Condensed Matter and Center for Sustainable Systems Design, Chair of Theoretical Chemistry, University of Paderborn, Paderborn33098, Germany

Corresponding author: Lena Simine lena.simine@mcgill.ca

[$]current affiliation: Department of Chemistry, New York University, New York City, New York, 10003, United States

ABSTRACT

Disordered molecular systems such as amorphous catalysts, organic thin films, electrolyte solutions, and water are at the cutting edge of computational exploration today. Traditional simulations of such systems at length-scales relevant to experiments in practice require a compromise between model accuracy and quality of sampling. To remedy the situation, we have developed an approach based on generative machine learning called the Morphological Autoregressive Protocol (MAP) which provides computational access to mesoscale disordered molecular configurations at linear cost at generation for materials in which structural correlations decay sufficiently rapidly. The algorithm is implemented using an augmented PixelCNN deep learning architecture that we previously demonstrated produces excellent results in 2 dimensions (2D) for mono-elemental molecular systems. Here, we extend our implementation to multi-elemental 3D and demonstrate performance using water as our test system in two scenarios: 1. liquid water, and 2. a sample conditioned on the presence of a rare motif. We trained the model on small-scale samples of liquid water produced using path-integral molecular dynamics simulation including nuclear quantum effects under ambient conditions. MAP-generated water configurations are shown to accurately reproduce the properties of the training set and to produce stable trajectories when used as initial conditions in classical and quantum dynamical simulations. We expect our approach to perform equally well on other disordered molecular systems while offering unique advantages in situations when the disorder is quenched rather than equilibrated.

INTRODUCTION

There are numerous examples of amorphous or strongly disordered systems that gave rise to decades of fruitful research or hold a terrific potential for future applications: amorphous silica [1], water[2], amorphous catalysts[3], amorphous graphene[4] to name a few. Computational exploration of such systems is hindered by the need to simulate large numbers of particles in difficult-to-sample free energy landscapes. The steep rise in simulation cost with system size is typically referred to as the 'exponential wall' or the 'curse of dimensionality' and it is due to the fact that the configuration space volume and the necessary sampling increase exponentially with number of particles. State of the art simulation methods approach the problem of bridging the nano-to-meso gap through the development of enhanced-sampling techniques[5–7], coarse-grained potentials[8–10] and machine-learning force fields[11,12]. They increase to some degree the range of systems that may be successfully modeled. Our approach seeks to exist in this space, and it effectively circumvents the dimensionality curse for the limited (although quite broad) class of strongly disordered systems.

Inspired by the successes of applying machine learning to sampling challenges, we have recently introduced an approach to simulating strongly disordered matter as an autoregressive protocol based on deep learning called the Morphological Autoregressive Protocol (MAP) [13,14]. It takes advantage of the locality of structural correlations in amorphous or strongly disordered materials and extrapolates molecular configurations from small-scale samples to arbitrary sizes at linear cost, see Figure 1 for a conceptual summary. Previously, we have implemented the MAP for the case of 2D mono-elemental materials using a deep generative model called gated PixelCNN[15,16] and applied it to simulations of quantum dot aggregates, and to a recently discovered topologically distinct amorphous variant of graphene[13,17,18]. Here, we extend it to multi-elemental 3D disordered structures.

The idea behind the MAP is that for materials with finite correlation length, a neural network trained on small, order of the structural correlation-length samples (Figure 1(A)), learns the conditional probability of presence/identity of atoms at a point in space given the 'context' of surrounding atoms (Figure 1(B)) and it can be used to extrapolate structures (Figure 1(C)) to any desired size at linear cost with respect to the number of the generated grid-points. If needed, the samples may reach the mesoscale, and if needed, the resolution may remain atomistic. In practice, we estimate the correlation length $L_c$ using a radial distribution function (RDF)

$$g(r) = \frac{1}{N} \langle \sum_{i \neq j}^{N} \delta(r - r_{ij}) \rangle$$

where $\delta$ is the Dirac delta function, $r$ the radius, $r_{ij}$ the distance of particle $j$ from the reference particle $i$, and $N$ the total number of particles. Once the probability conditional on the context with the size of $L_c$ is known, it is then sampled in an autoregressive fashion with newly generated points becoming input for further generation. Using the MAP, the molecular structures are generated in a directional manner one grid-point at a time and the required input 'molecular context' is limited to the space directly preceding the generated grid-point. We found that 'molecular context' truncated in this fashion is sufficient for successful extrapolation in the case of 2D amorphous materials[13] and here we show that it is indeed sufficient for the 3D case as well.

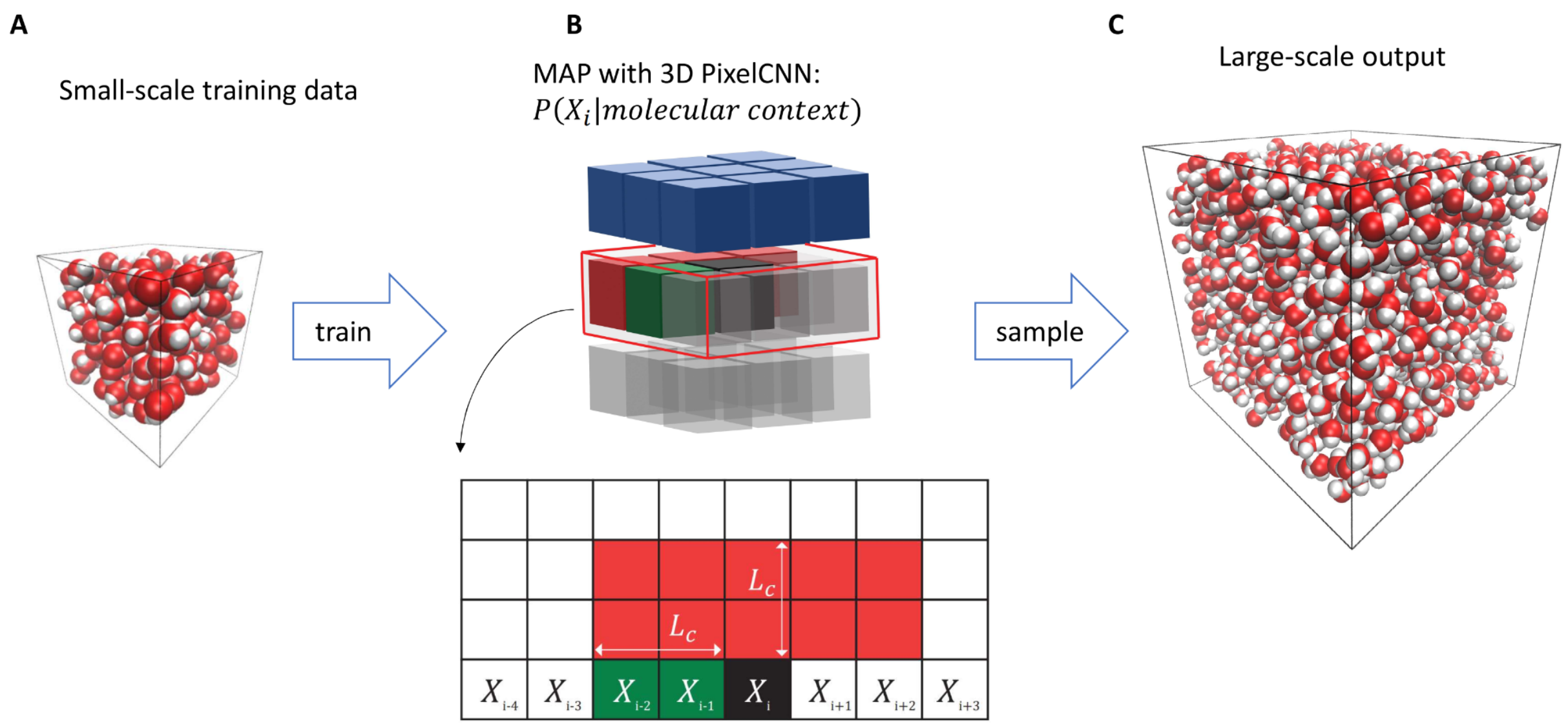

**Figure 1** Workflow diagram for Morphological Autoregressive Protocol (MAP) in 3D. (A) Training samples (size on the order of the correlation length $L_c$). (B) Schematic representation the outcome of training the 3D PixelCNN model: a conditional probability of the grid-point $X_i$ given the context of size $L_c$: $P(X_i|molecular\ context)$. C. The outcome of the generation process: an extrapolated molecular configuration with dimensions that exceed the dimensions of the training samples.

Finally, it is important to discuss whether the generated structures are physical. We showed in Ref.[14] that 1. MAP-generated structures can be thought of as converging Markov chains that are guaranteed to realize a unique steady-state distribution (the distribution of the small-scale samples used in training) and 2. The extrapolation scheme in the MAP is physically motivated by the decaying correlation length in amorphous or strongly-disordered materials and, therefore, the training and the generation process can be systematically improved/converged through the tuning of the relevant hyperparameters and the design of the training set. The easiest to tune and the most effective generation hyperparameters are the sampling temperature and the size of the 'burned' region – the part of the early generated sample that is discarded due to lack of equilibration. As far as the design of the training set goes, the training samples need to be sufficiently large to allow structural correlations to decay, and the data needs to be appropriately augmented (see Methodology section) in order to include symmetries inherent to the system during training. Competing approaches that resemble the MAP have recently emerged based on generative adversarial networks[19–21], but to the best of our knowledge so far the physicality of the generated samples has not been systematically explored.

In this paper we present an implementation of the MAP for the simulation of 3D multi-element atomistic molecular structures. To demonstrate the performance of multi-elemental MAP in 3D we use liquid water as model system. Water has various anomalous properties that have been of great interest for more than five decades[22,23]. Here we first verify that our model successfully generates liquid water in ambient conditions. We then proceed to generate structures that are conditioned on the presence of a rare motif found in the training dataset and we demonstrate convergence of the properties of new configurations with the number of generative iterations. We propose that this novel approach may be useful for studying challenging scenarios like ice nucleation etc. Finally, we seed molecular dynamics simulations and confirm that resulting trajectories are stable and physically sound. The paper is organized as follows: in Methodology section we present the details of the MAP and its multi-elemental implementation in 3D and the technical details of the molecular dynamics simulations, Results and Discussion section discusses the MAP performance on liquid water dataset and the results of molecular dynamics simulations seeded with configurations generated by MAP, Conclusions provide an executive summary and outlook.

## METHODOLOGY

*1.The Morphological Autoregression Protocol (MAP).*

In the MAP approach, the space is discretized, and the molecular structure is viewed as a sequence {X}. The probability that grid point $i$ is occupied by a molecular fragment class $\theta$ (atomistic or otherwise) is generically given by

Equation 1: $P(X_i = \theta|\{X_{j<i}\}) = N[c_{b,\theta} + F_\theta(\{X_{j<i}\})]$

with $\{\theta \in \mathbb{Z}^+|\theta \leq N_c\}$ where $N_c$ is the total number of classes, $c_{b,\theta}$ is a linear bias, $F_\theta(\{X\})$ is a function that captures the correlations between $X_i$ and the surrounding structure $\{X_{j<i}\}$ and $N[y]$ signifies the soft-max class-wise normalization function that translates $\boldsymbol{y}_{\theta,i} \doteq c_b + F_\theta(\{X_{j<i}\})$, into the probabilities for atomic/molecular features expected at grid-point $i$ (with $\beta$ = 1):

Equation 2: $N[y_{\theta,i}] = \frac{e^{\beta y_{\theta,i}}}{\sum_\theta e^{\beta y_{\theta,i}}}$.

With index $i$ running over multiple dimensions, one can generate 1, 2 or 3 dimensional sequences corresponding to 1D, 2D, and 3D materials. Uniquely, finite morphological correlation lengths in disordered materials allow us to truncate the molecular context at some finite correlation length $L_c$ (Figure 1(B)) without loss of accuracy, in other words:

Equation 3: $lim_{i\to\infty} F_\theta(\{X_{j<i}\}) = F_\theta(\{X_j \in L_c\})$.

Thereby, the probability for a sequence element to be of class $\theta$ is given by:

Equation 4: $P(X_i = \theta|\{X_{j<i}\}) = N[c_{b,\theta} + F_\theta(\{X_j \in L_c\})]$:
which are normalized, nonzero probabilities for each class, see Equation 2. Sampling this conditional probability starting from either empty space or from some initial conditions will generate the sequence {X} element by element with newly generated elements fed back into the conditional probability as part of the molecular context $\{X_j \in L_c\}$.

The correlation function $F_\theta$ in Equation 4 may have different forms depending on the system that is being studied. For simple systems such as dilute gases or fluids, one can derive $F_\theta$ by incorporating pairwise correlations of particles, but this is not a trivial task for dense fluids and similarly for the amorphous materials where one needs to take into account correlations of two pairs, three pairs, one 3-body cluster, one 3-body and one pair, etc., up to large clusters of particles. The number of these terms explodes exponentially with correlation length $L_c$ and size of the aggregates. Therefore, we resort to fitting $F_\theta$ as accurately as possible using data. In the next section, we describe the PixelCNN architecture which is known for its effectiveness in learning complex multi-dimensional datasets.

*2.PixelCNN:*

PixelCNN is a variant of convolutional neural networks which explicitly outputs the conditional probability distribution for all the grid-points in 3D or voxels in an image, {X}, given the surrounding voxels:

Equation 7: $P(\{X\}) = \prod_i^{n_r \cdot n_c \cdot n_l} P(X_i|\{X_{j<i}\})$.

Here the set {X} represents the training data in the sequential form with $n_r$, $n_c$ and $n_l$ corresponding to the number of rows, columns, layers of the 3D input data. To ensure the receptive field of the convolutions around each target voxel, here with size of 3x3x3, only includes the voxels on which its probability is conditioned (thus, avoiding seeing the future voxels/context) a mask is added to each convolution, see Figure 2(A).

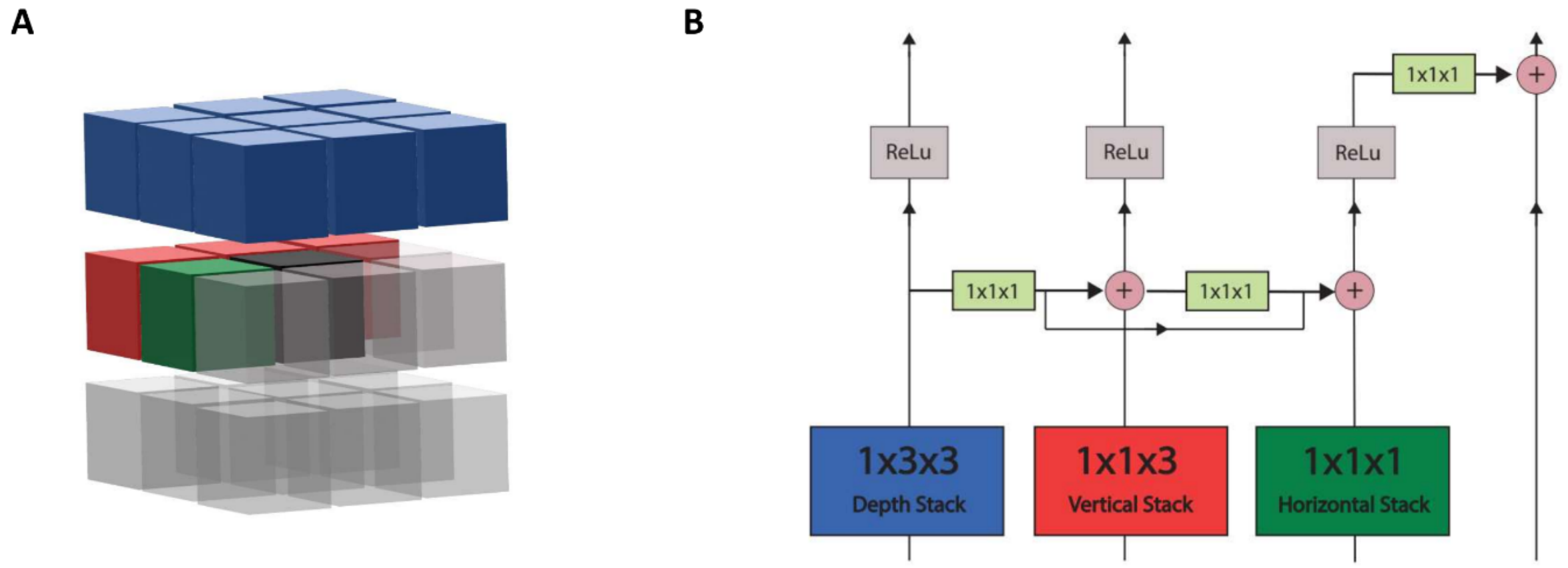

**Figure 2** Details of the 3D-MAP-PixelCNN architecture. (A) Masked kernel (3x3x3) centered on a target voxel (black), the voxels prior to the target voxel are shown in blue, red and green corresponding to the prior voxels being convoluted in depth, vertical, horizontal stacks respectively. (B) depicts the architecture of PixelCNN in 3D, with stacks working independently to overcome the blind spot problem.

However, the use of 'masked' convolutions leads to loss of information, this artifact is known as blind spot. The elimination of the blind spot is achieved by use of two masks - vertical and horizontal[15] in 2D. We extend this approach to 3 dimensions by splitting the masked 3x3x3 convolution into a 1x1x1 (horizontal) and 1x1x3 (vertical) convolution and 1x3x3 (depth) convolution. Thus, these convolutions are done in 3 separated stacks demonstrated in Figure 2(B). Note that the information from depth stack is flowed to vertical and horizontal, and from vertical to horizontal in a unidirectional manner by 1x1x1 connections. Depth stack should not have access to any information the vertical has, and similarly vertical stack is blind to the information accessible to the horizontal stack, otherwise they would access the voxels that they should not see.

We used Rectified Linear Units (ReLU) activations for each layer. As in Ref [15], after the first layer we add a residual connection from the ReLU unit in horizontal to the output of the next one. Moreover, the final convolution is accompanied by dropout, a well-known regularization method[25]. In the generation phase, the softmax operation is followed by multinomial sampling using the normalized voxels probabilities, to predict the voxel value. This step is the implementation of Equation 4 described in the MAP protocol. The predictions for each voxel are compared with the associated voxel value in training sample by using a cross-entropy loss function per-voxel defined as:

$$L_{CE} = -\sum_{\theta} t_{\theta} \log P(X_i = \theta)$$

where $t_{\theta}$ is the one-hot encoded truth value of each voxel, and $P(X_i = \theta)$ is the prediction made by Equation 4. Finally, a backpropagation algorithm is performed by using the computed loss function for every batch of training sample to update the parameters of the model.

*3. Molecular dynamics simulations: training set and validation*

The training set was generated using a quantum mechanical path integrals molecular dynamics (PIMD) trajectory (i.e., including nuclear quantum effects) sampled at 1.25 ps rate to make a total of 1000 decorrelated configurations of 216 water molecules[31]. For the validation of the generated samples, the classical molecular dynamics (MD) and PIMD simulations were performed at 300 K MAP generated samples with a total of 343 water molecules. Periodic boundary conditions were applied using the minimum image convention. First a NPT equilibration was done for 125 ps, followed by a NVT run for 500 ps at 1 bar using an Andersen thermostat and an anisotropic Berendsen barostat, respectively[26,27]. For all calculations the q-TIP4P/05 water potential of Habershon et al.[28] has been used. Short-range interactions were truncated at 10 Angstrom and Ewald summation was employed to calculate the long-range electrostatic interactions. We used discrete time steps of 0.5 fs for the intermolecular and 0.1 fs for the intramolecular forces[29]. For the PIMD calculations the ring polymer contraction scheme with a cut-off value of σ = 5 Angstrom was employed to reduce the electrostatic potential energy and force evaluations to a single Ewald sum, speeding up the calculations[30]. 32 beads were used, contracting to a centroid for the computational expensive part of the electrostatic interactions. Ensemble averages were calculated over the whole trajectory length of 500 ps.

RESULTS AND DISCUSSION

*1. Liquid Water:* For training on liquid water structures, 1000 samples were generated by PIMD simulations (see Methods for details). We augmented this dataset with samples rotated 90 and 180 degrees in each direction to achieve a training set containing 7000 samples with volume of 80x80x80 voxels. Each voxel is equal to 0.25Å$^3$ introducing an average distortion in atom positions of 4 % of O-H covalent bond length upon discretization of space. Finally, to reduce the memory usage, we subsampled each configuration with 8 equally sized and non-overlapping samples, this leads to 56000 40x40x40voxels samples. This data set was split 80:20 into training and test data sets. We chose model architecture with 20 layers each having 20 convolutional kernels. Stochastic gradient descent (SGD) with momentum of 0.9 was used, and the training was done using cross entropy loss function and stopped after 493 epochs with the batch size set to 1 and the learning rate to 0.01. A dropout with probability of 0.21 was used after the final layer which randomly sets to zero one of the elements of the features tensor.

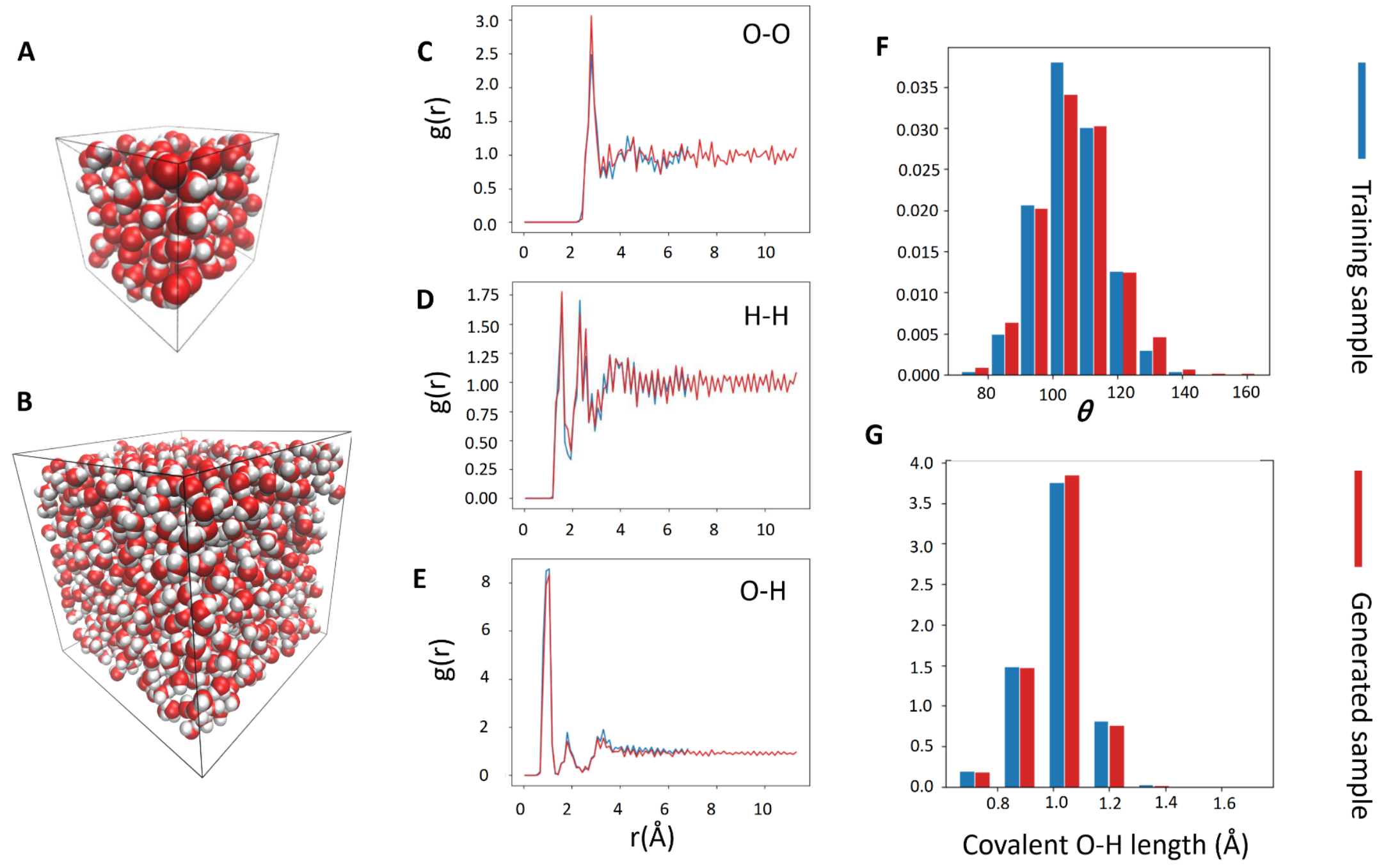

**Figure 3** Performance of the MAP on a model system (ambient liquid water). (A) A sample of liquid water from the training set. (B) An extrapolated larger MAP-generated sample with sides of the simulation box doubled relative to the training set. Comparison of the properties of the training set (blue) and the properties of MAP-generated samples (red): Panels (C), (D) and (E) give the radial pair correlations for O-O, H-H, and O-H, and (F), (G) show distributions of H-O-H bond angles and covalent O-H bond lengths.

Figure 3 shows the results of MAP simulations of ambient water and compares characteristics of the training samples to those generated by the MAP. For this comparison we have generated one sample with dimensions of 160x160x160 voxels or 4x4x4 $nm^3$, i.e., each side is twice as large as the training data samples, for an illustration, see Figure 3(B). The overall comparison demonstrates strong performance with excellent agreement between the training and the generated samples. We use element-resolved radial distribution function to evaluate agreement: see Figure 3(C) for Oxygen-Oxygen RDF, Figure 3(D) for H-H RDF, and Figure 3(E) for O-H RDF. The radial distribution function g(r) function was calculated by counting the number of particles contained in a volume within the spheres of radii $r$ and $r + dr$ away from a reference particle. As the training samples and outputs are in the discretized space, $dr$ was chosen to be half of voxel side (0.125Å). Note that this discretization gives rise to the noisy RDFs in Figure 3. To account for normalization, the counts in previous step were divided by $\rho\ 4\pi r^2 dr$, where $\rho$ is the number density. The evidence that the model captures the water bond angle and bond lengths very well is presented in Figures 3(F) and (G). Finally, we note that our model is prone to minor mistakes leaving oxygens and hydrogens 'orphan' in rare cases. The number of these defects was in total 88 among 1923 generated water molecules with the majority found at the boundaries. This is expected since we fix the number of voxels in the generated sample rather than the number of molecules and molecules initiated at the edges do not get completed. These defects can be removed with a simple post-processing step.

2. *Configurations conditioned on presence of a rare motif:* Here, we explore the potential of MAP to generate configurations of liquid water in the presence of a rare motif. To obtain the rare motif, we calculated the density of water molecules in a probe volume of 25x25x25 voxels within the training set. While the local structures with a density of 0.996 g/cm³ are dominant, infrequent low-density structures with a density of 0.614 g/cm³ have been discovered. These low-density motifs have a probability of about 4.5 % among the entire existing water molecules structures within the probe volume. We chose a specific training sample where the low-density motif is found exactly at the center of the simulation box. Our aim is to generate a configuration of the same size as this training sample, with the motif placed at the center of the generated sample and water is generated around it. We generate such a configuration as follows: (i) initialize an empty sample with only the motif configuration present and populate the surrounding grid-points with our model, Figure 4(A); (ii) we rotate the generated sample counterclockwise along the z (blue axis) and clockwise along the x direction (red axis); (iii) we now populate the grid-points that remained empty after step (i), see Figure 4(B). Consequently, we have a sample with a desired motif in the center of the box and water molecules around it sampled from a distribution that was conditioned on the presence of the motif, Figure 4(C). Given that the MAP generation is a Markov chain[14], it is anticipated that several iterations of MAP generation would be required for convergence to a steady state distribution. To perform iterative generation, we repeat the generation protocol (step(ii) to (iii)) but erase those grid points that will be generated at the next iteration step and generate them anew.

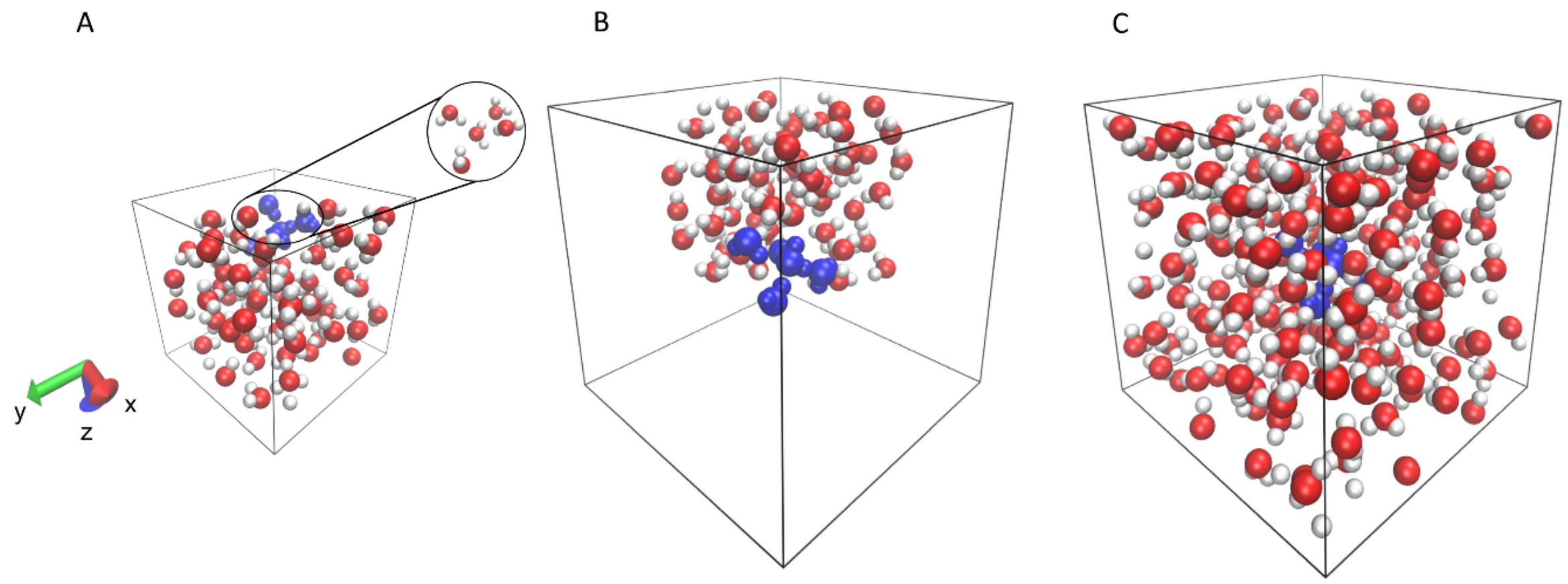

**Figure 4** Visual summary of conditional generation using the MAP. (A) Step (i) in conditional generation: generated water molecules around a rare motif (blue molecules) seeded in the corner of the box. (B) Step (ii) in conditional generation: rotated configuration with empty space to be filled in Step (iii). (C) Step (iii) in conditional generation: the rest of the simulation box is filled in. As a result, the motif is placed in the center of the box surrounded with water molecules.

In order to investigate the convergence of the motif-conditioned sample as a function of generative iteration we quantify the similarity between the generated samples and the training sample from which the motif is drawn. To do so, we use the local bond order parameters $q_4$ and $q_6$ applied to the oxygen atoms. These are based on the Steinhardt bond-order parameters[32,33] and have been widely employed to understand structural ordering in systems like water and amorphous solids[34]. These order parameters are defined as follows:

$$q_l(i) = \left( \frac{4\pi}{2l+1} \sum_{m=-l}^{l} |q_{lm}(i)|^2 \right)^{\frac{1}{2}}$$

where,

$$q_{lm}(i) = \frac{1}{N_{neigh}(i)} \sum_{j=1}^{N_{neigh}(i)} Y_{lm}(\boldsymbol{r_{ij}})$$

in which $N_{neigh}(i)$ is the number of neighbors of particle $i$, $\boldsymbol{r_{ij}}$ is the vector connecting particle $i$ with its neighbors $j$ within a cut of distance of 3.5 Å and $Y_{lm}$ are spherical harmonics of order $l$ and $m$.

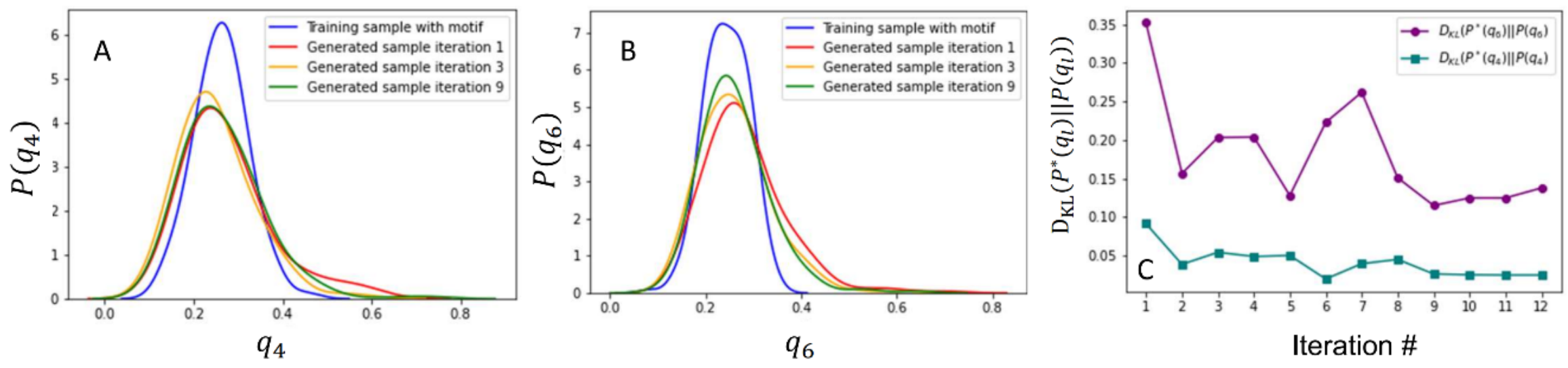

**Figure 5** Probability density functions for local order parameters $q_4$ (A) and $q_6$ (B) in a reference sample containing a rare low-density motif and in MAP-generated samples conditioned on that motif shown as a progression of generative iterations. Panel (C) demonstrates the KL divergence ($D_{KL}$) between the distributions of $q_4$ and $q_6$ from the generated samples and the reference. Convergence with increase in number of generative iterations is observed towards a steady state as well as (weakly) towards the reference state.

We present the probability density functions for $q_4$ and $q_6$ in Figure 5(A) and (B) respectively for the different iterations of generated samples containing a rare motif (red, orange, green) alongside the reference sample (blue). Figure 5(C) shows the Kullback-Leibler (KL) divergence between local order parameters' distributions in the reference sample $P^*(q_l)$ and in the generated samples $P(q_l)$ as a function of number of generative iterations. It is defined as: $D_{KL}(P^*||P) = \int P^*(q_l) \log \frac{P^*(q_l)}{P(q_l)} dq_l$ with the index $l \in 4{,}6$. The KL divergence measures the discrepancy between two distributions, and we use it to track convergence of the generated samples. It is clear from the trend in Figure 5(C) that several iterations are necessary for producing converged configurations. In our experience, $q_4$ tends to converge more readily than $q_6$ since the latter incorporates more intricate angular interactions among adjacent particles, and it is more sensitive to positions of the atoms. Note that the order parameters are not expected to match the reference exactly (and the KL is not expected to go to zero) as we do not expect the exact same configuration to be generated by the model. Nonetheless, it is reassuring to observe a weak trend towards the reference in these data.

*3. Molecular dynamics simulations*

In order to put the MAP-generated configurations to a practical test we used them to initialize several molecular dynamics trajectories and kept track of the physicality of observables. To this end, we have generated a sample with dimensions of 2.4x2.4x2.4 $nm^3$ that includes 343 water molecules and used it as initial configuration in a set of classical MD and quantum mechanical PIMD simulations including nuclear quantum effects. The results are summarized in Figure 6 and Table 1.

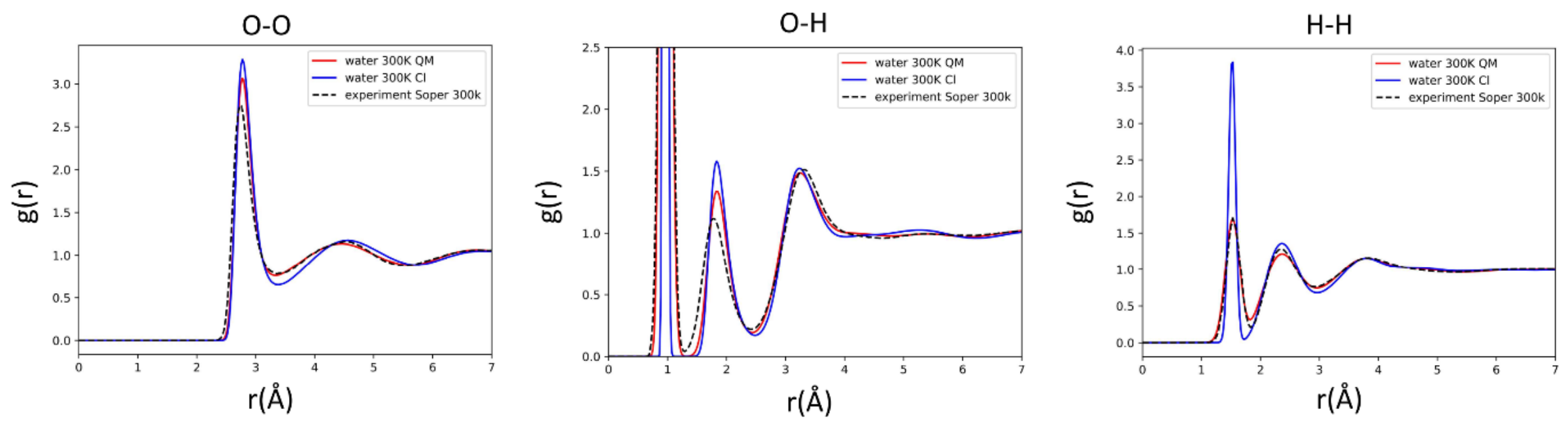

**Figure 6** Oxygen-oxygen (left), oxygen-hydrogen (middle), and hydrogen-hydrogen (right) Radial Distributions Function (RDF) for water at 300 K with NQE (red), classical water at 300 K (blue) and experimental water at 300 K (dotted). All calculated RDFs of the simulations are computed from 10000 steps from quantum mechanical PIMD including nuclear quantum effects (NQE) with 250 ps length and 20000 steps of classical path integrals molecular dynamics (PIMD) calculation with a total length of 500 ps.

Overall, the water at ambient temperature shows the typical behavior we would expect, similar to experimental and theoretical studies done before[35–37]. The O-O, O-H and H-H radial distribution functions (RDFs) for the classical and the quantum simulations are shown in Figure 6, and they show very similar structures. Table 1 summarizes key observables from the two simulations which are in agreement with generally accepted values[38]. This confirms that MAP-generated starting structures give rise to healthy dynamical simulations of water.

| | Density $[g/cm^3]$ | Average O-H bond length [Å] | Average bond angle [degrees] | Average dipole [Debye] |
|---|---|---|---|---|
| Water 300 K MD | 1.0003 | 0.9623 | 104.825 | 2.308 |
| Water 300 K PIMD | 0.9958 | 0.9779 | 104.710 | 2.347 |
| Experimental water 300 K | 0.997 | 0.97 | 105.100 | 2.9 |

Table 1: Density and average observables over the NPT followed by NVT runs for classical and quantum molecular calculations. Experimental data taken from Ref. [35].

## CONCLUSIONS

In this paper we have extended to 3D and multi-elemental datasets the morphological autoregressive protocol (MAP), a recently constructed computational approach to generating strongly disordered morphologies at arbitrary scales based on small-scale training samples. By construction, the MAP can be applied to any molecular dataset with sufficiently quickly decaying structural correlation function. Here, we have demonstrated stable performance using water as a model system and highlighted some of the aspects in which the MAP can benefit computational exploration, for example 1. In the case of water, NQE are naturally included in the generated morphologies by virtue of their inclusion in the training data. This may be important for low temperature simulations where water is a quantum liquid; 2. MAP-generated initial conditions may be used to bypass expensive sampling using (PI)MD simulations by motif-conditioned generation. The selection of the motif is influenced by the specific problem being addressed and the presence of the required motif in the training data. Beyond water, we expect this deep learning protocol to be of use in a wide range of simulations of strongly disordered materials by providing cost-effective and direct ways of probing the morphological complexity, gaining access to important rare configurations, and ultimately bridging the gap between the nano- and the meso- scale simulations of molecular systems.